\documentclass[iop]{emulateapj}

\usepackage{apjfonts}
\usepackage{pbox}
\usepackage{bm}
\usepackage{CJK}

\shorttitle{Red clump stars in LAMOST and APOGEE}
\shortauthors{Ting et al.}

\begin{document}

\begin{CJK*}{UTF8}{gbsn}
\title{A large and pristine sample of standard candles across the Milky Way:\\ $\sim 100,$000 red clump stars with 3\% contamination}
\author{Yuan-Sen Ting (丁源森)\altaffilmark{1,2,3.4}, Keith Hawkins\altaffilmark{5}, Hans-Walter Rix\altaffilmark{6}}
\altaffiltext{1}{Institute for Advanced Study, Princeton, NJ 08540, USA}
\altaffiltext{2}{Department of Astrophysical Sciences, Princeton University, Princeton, NJ 08544, USA}
\altaffiltext{3}{Observatories of the Carnegie Institution of Washington, 813 Santa Barbara Street, Pasadena, CA 91101, USA}
\altaffiltext{4}{Research School of Astronomy and Astrophysics, Australian National University, Cotter Road, ACT 2611, Canberra, Australia}
\altaffiltext{5}{Department of Astronomy, Columbia University, 550 W 120th St, New York, NY 10027, USA}
\altaffiltext{6}{Max Planck Institute for Astronomy, K\"onigstuhl 17, D-69117 Heidelberg, Germany}
\slugcomment{Submitted to ApJL}

%
%
%
%
%
%
\begin{abstract}
Core helium-burning red clump (RC) stars are excellent standard candles in the Milky Way. These stars may have more precise distance estimates from spectrophotometry than from Gaia parallaxes beyond 3$\,$kpc. However, RC stars have $T_{\rm eff}$ and $\log g$ very similar to some red giant branch (RGB) stars. Especially for low-resolution spectroscopic studies where $T_{\rm eff}$, $\log g$, and [Fe/H] can only be estimated with limited precision, separating RC stars from RGB through established method can incur $\sim 20\%$ contamination. Recently, \citet{haw18} demonstrated that the additional information in single-epoch spectra, such as the C/N ratio, can be exploited to cleanly differentiate RC and RGB stars. In this second paper of the series, we establish a data-driven mapping from spectral flux space to independently determined asteroseismic parameters, the frequency and the period spacing. From this, we identify 210$,$371 RC stars from the publicly available LAMOST DR3 and APOGEE DR14 data, with $\sim 9\%$ of contamination. We provide an RC sample of 92$,$249 stars with a contamination of only $\sim 3\%$, by restricting the combined analysis to LAMOST stars with S/N$_{\rm pix}\ge 75$. This demonstrates that high-S/N, low-resolution spectra covering a broad wavelength range can identify RC samples at least as pristine as their high-resolution counterparts. As coming and ongoing surveys such as TESS, DESI, and LAMOST will continue to improve the overlapping training spectroscopic-asteroseismic sample, the method presented in this study provides an efficient and straightforward way to derive a vast yet pristine RC stars to reveal the 3D structure of the Milky Way.
\end{abstract}


%
%
%
%
%
%

\section{Introduction}
\label{sec:intro}

Low-mass stars will evolve off the main-sequence at the end of their core hydrogen burning phase: during the red giant branch (RGB) ascent the star has an inert helium core surrounded by a hydrogen burning shell \citep{ibe68}; they will then go through the helium flash (for M $\geq$ 0.8~M$_{\sun}$) and quickly descend in the $T_{\rm eff}$~-~$\log g$ diagram to reach the core helium burning phase called the red clump (RC). 

For mapping the Galaxy, RC stars are exciting and highly sought-after tracers, because their brightness and color are well-constrained given their metallicity and age \citep[e.g.,][]{sta98,gir16,haw17}. Their tight, well-defined position in color--absolute magnitude space makes them exceptional standard candels, when combined with good photometry. This then enables precise 3D mapping of stars and gas in the Milky Way. Gaia will, of course, provide parallax-based distance estimates for RC stars. For example, for unreddened stars the expected distance precision will be $\sim 10\%$ at $\sim 3~{\rm kpc}$ for Gaia DR2 data. But the distance precision based on parallaxes will rapidly deteriorate beyond 3$\,$kpc and for stars in the Galactic plane that are highly reddened.. On the other hand, using RC stars, we can achieve high precision distance estimates, with precise photometry, up to $\sim 10$ kpc, with a distance errors of $\sim 6\%$ \citep{bov14,haw17}. Therefore, ``clean'' and extensive samples of RC stars is a key to unravel the structure of the distant Milky Way beyond the Galactic neighborhood.

Asteroseismic parameters of a giant star -- in particular the large frequency separation between $p$-modes, $\Delta \nu$, and the period spacing of the mixed $g$ and $p$ modes, $\Delta P$ -- provide a remarkably clean separation of RC and RGB stars \citep[e.g.,][]{bed11}. Of course, asteroseismology probes the stellar interior, and hence plausibly can diagnose core helium burning in a star, while spectroscopy at face value only probes surface properties. Therefore, clean spectroscopic classification of RC stars has proven to be challenging, because RC stars show similar $T_{\rm eff}$ and $\log g$ to part of the RGB phase. The established spectroscopic classification approach is to classify a star as RC or RGB on the basis of $T_{\rm eff}-\log g-$[Fe/H] and their color-magnitude, when compared to theoretical expectation from isochrones \citep[e.g.,][]{bov14}. Precise classification with this approach requires a precision of $T_{\rm eff}$ and $\log g$ to be much better than 100$\,$K and 0.1$\,$dex, respectively, a condition hard to meet especially for low-resolution spectroscopy. Consequently, this approach typically incurs a contamination rate of $\sim 20\%$ for low-resolution spectra \citep{wan15}. 

\begin{figure*}
\centering
\includegraphics[width=1.0\textwidth]{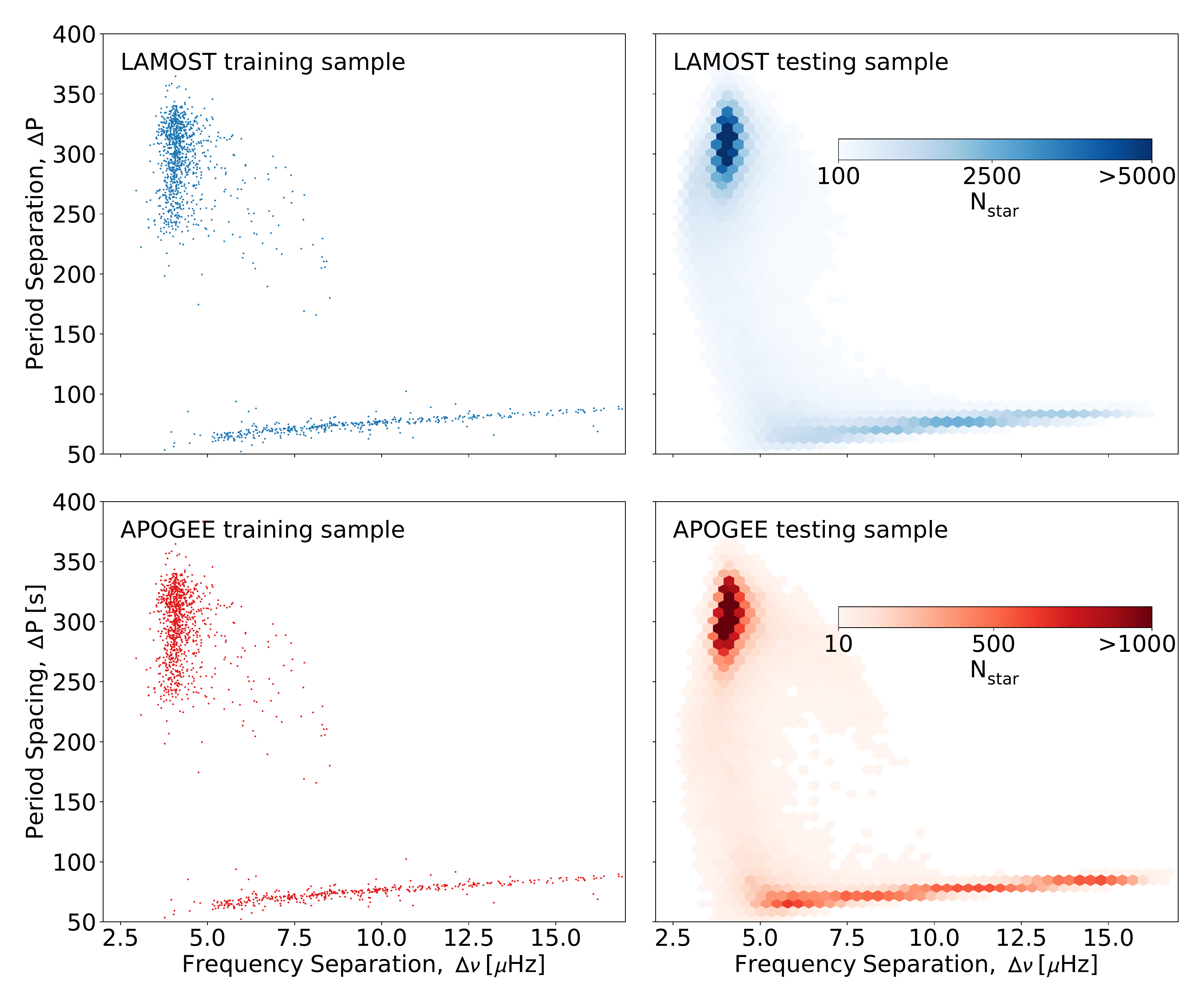}
\caption{Single-epoch spectra can infer precise asteroseismic parameters. The left panels illustrate the asteroseismic values $\Delta \nu$ and $\Delta P$ from \cite{vra16} for the training sample, used to construct the mapping from spectra to asteroseismic parameters. Plotted on the right panels are the results when applying this mapping to all APOGEE and LAMOST spectra that have similar spectroscopic stellar parameters as the training set. Inferring asteroseismic parameters from spectra yields a cleaner separation for RC and RGB stars -- there is a distinct bimodal distribution in $\Delta P$. The RC stars are centered at around $\Delta P = 300\,$s and the RGB stars around $\Delta P = 70\,$s.}
\label{fig1}
\end{figure*}

But this limitation can be overcome when considering {\em full} spectral information, as illustrated by \citet{haw18}, where they showed that photospheric abundances in APOGEE spectra must reflect the interior structure of the stars via the efficacy of extra-mixing on the upper RGB \citep{mar08,mas15,mas17a,mas17b} and can therefore be used to infer asteroseismic parameters and discriminate RC from RGB stars. \citet{tin17b} demonstrated that even at low-resolution (R=2000), spectra do contain spectral information of many abundances, going beyond the basic stellar parameters. Combining these studies implies that with a suitable spectro-asteroseismic training data, a data-driven empirical mapping that relates low-resolution spectra to asteroseismic parameters can be established, and provides a much clean separation of the RC and RGB stars, which is the focus of this study.

\begin{figure*}
\centering
\includegraphics[width=1.0\textwidth]{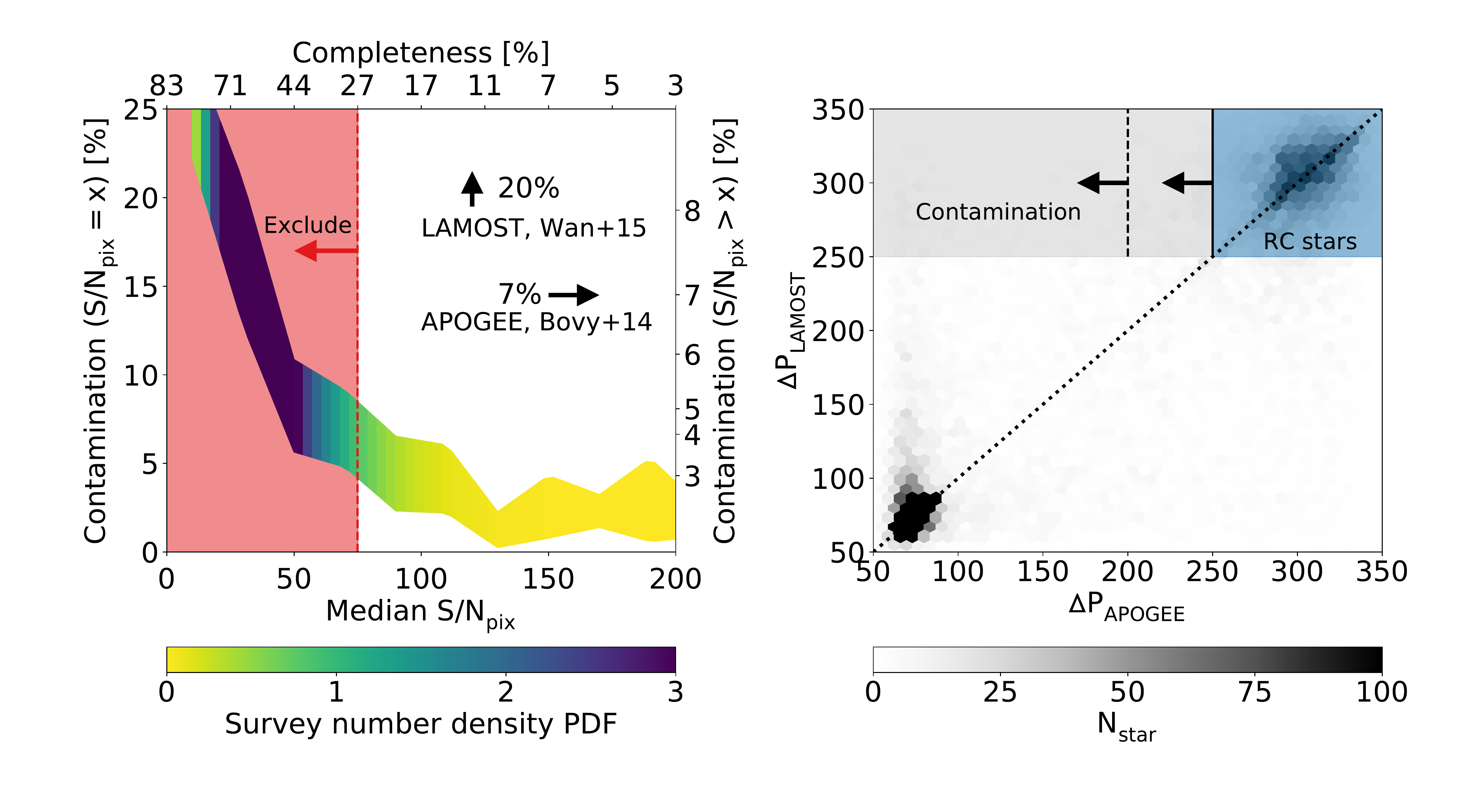}
\caption{Purity of the RC sample derived in this study. The right panel shows the $\Delta P$ spectroscopic estimates of the 14$,$442 overlapping test stars in APOGEE and LAMOST, which we will adopt as the cross-validating set to determine the contamination rate. The training of the APOGEE and LAMOST empirical relation is independent. The agreement demonstrates that the inferred asteroseismic parameters from spectra are mostly robust. The left panel further quantifies this robustness. To be conservative, we only include stars with $\Delta P > 250\,$s to be RC stars in our catalog. Shown is the contamination rate as a function of S/N$_{\rm pix}$ of the LAMOST spectra. The dual $y$-axes show the contamination rate at a specific S/N$_{\rm pix}$ on the left and the integrated contamination rate for all stars beyond the S/N$_{\rm pix}$ selection criterion on the right. The upper limit is defined as the fraction of stars that have $\Delta P_{\rm LAMOST} > 250\,$s but with $\Delta P_{\rm APOGEE} < 250\,$s, and the lower limit with $\Delta P_{\rm APOGEE} < 200\,$s. The completeness, defined as the fraction of the number of stars having $\Delta P_{\rm LAMOST} > 250\,$s to the total number of stars having $\Delta P_{\rm LAMOST} > 200\,$s, is illustrated in the top $x$-axis. Color-coding indicates the normalized number density of the full LAMOST test set in this study. Even LAMOST spectra are low-resolution, the method presented can achieve a pristine separation of RC and RGB stars. The primary limiting factor is the S/N of the spectra but not the method itself. For this study, our main catalog is restricted to LAMOST stars with S/N$_{\rm pix}> 75$ to guarantee the purity of the sample, such that the integrated contamination is only $\sim 3\%$. We note that this adversely affects the completeness as shown in the top $x$-axis because most LAMOST stars have low S/N. But we will also release our estimates for all stars.}
\label{fig2}
\end{figure*}

%
%
%
%
%
%

\section{Method: Direct Spectral Separation of Red Clump and Red Giant Branch Stars}

In this paper we set out to establish a data-driven mapping from spectra to asteroseismic parameters, using an extensive sample of stars with asteroseismology from Kepler and spectra from APOGEE \citep{maj17} and LAMOST \citep{xia17b}. The spectra serve as input to predict the asteroseismic parameters as output. Once such a mapping is optimized, it can then be applied to all spectra to predict their asteroseismic values from such spectra. We will apply this approach to two large-scale spectroscopic surveys that span different resolutions and wavelength coverage as a sample application of this method. We will adopt the high-resolution ($R=22$,$500$) infrared ($\lambda = 1500-1700\,$nm) APOGEE survey and the low-resolution ($R=1800$) optical ($\lambda =390-900\,$nm) LAMOST survey. We will cross-validate this method with the sample overlapping among these two surveys.

First, we need to construct a training set that has both spectra and asteroseismic values. Following \citet{haw18}, we adopt the asteroseismic sample from \citet{vra16}, which provides estimates for both the frequency separation ($\Delta \nu$) between adjacent acoustic $p$-modes and the period spacing of the mixed gravity $g$- and acoustic $p$-modes. We restrict ourselves to objects from the Vrard catalog that have asteroseismic values consistent with the SSI catalog\footnote{http://ceps.spacescience.org/asteroseismology.html} ($|\Delta \nu_{\rm SSI} - \Delta \nu_{\rm Vrard}| < 2~\mu$Hz). Cross-matching this sample with the full APOGEE DR14 and LAMOST DR3 catalogs yield samples of $2853$ and $1137$ stars in common, respectively. Here, we only consider stars that have median spectroscopic S/N$_{\rm pix} > 50$ to avoid noisy training data. For each of the two spectroscopic-asteroseismic sample we set aside $100$ stars as validation set for the training step; in particular, we terminate the iteration of the training step (below), when the models stops to improve the asteroseismic parameter prediction for this validation set.

We continuum normalize all APOGEE spectra following \citet{nes15a} in which a fourth order polynomial is fitted to a subset of pixels that have the least data-driven gradients. As for LAMOST, we continuum normalize the spectra following \citet{ho17a} and \citet{tin17b}, where a smoothed version of the spectrum with a kernel size of 50\AA$~$is adopted as the continuum.

Since the empirical relation that we will establish below is trained on a defined training set which spans a specific stellar parameter regime, it is crucial to only apply this relation to stars with similar stellar parameter, not extrapolating too far from the training values. To ensure this, we follow an approach developed in \citet{tin16b}. We adopt the $T_{\rm eff}, \log g$ and ${\rm [Fe/H]}$ values of the APOGEE-asteroseismic and LAMOST-asteroseismic training sets from LAMOST DR3 and APOGEE DR14, and construct 3D convex hulls based on these stellar parameter values. We will only attempt to estimate asteroseismic parameters for stars in APOGEE DR14 and LAMOST DR3 that are within these restricted minimum convex polygon which encompasses the training set. Imposing this selection criterion leaves us a total of 413$,$472 stars from LAMOST DR3 and 80$,$944 stars from APOGEE DR14.

\begin{figure*}
\centering
\includegraphics[width=1.0\textwidth]{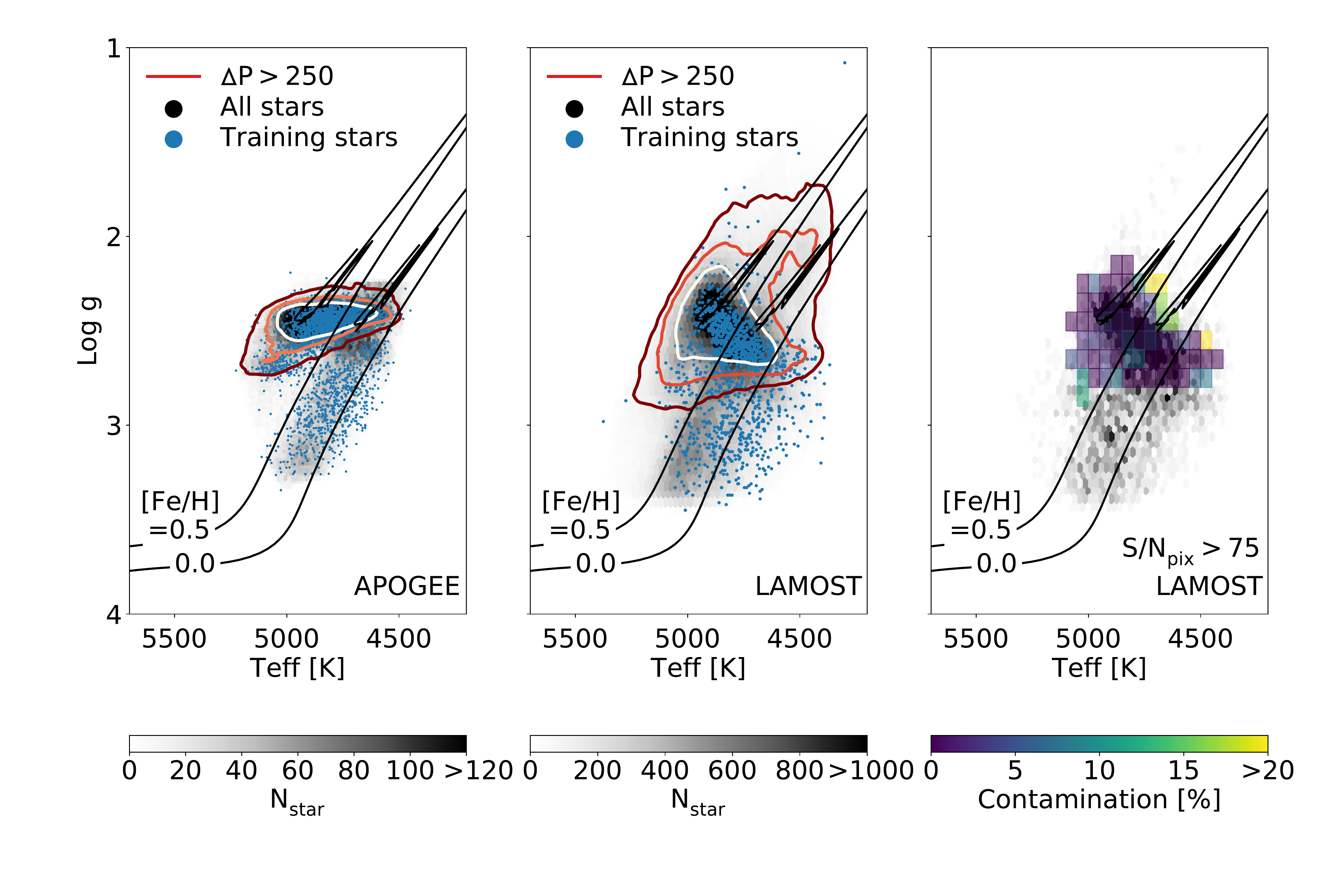}
\caption{The left and middle panels show the distribution of RC stars determined in this study in the $T_{\rm eff}-\log g$ diagram. We consider all stars with inferred $\Delta P > 250\,$s to be RC stars. Shown are the $T_{\rm eff} - \log g$ values from LAMOST DR3 and APOGEE DR14. The grey background demonstrates the number density of the test stars in this study. The blue symbols show the $T_{\rm eff}$ and $\log g$ of the training spectroscopic-asteroseismic sample, and the contours show the densities of the determined RC stars at 60, 80 and 95 percentiles, respectively. Overplotted in solid black lines are the MIST isochrones at 4 Gyr. The determined RC stars are located at the expected locus predicted by the MIST models, suggesting that the classification is robust. In the right panel, we illustrate the contamination rate for LAMOST stars with S/N$_{\rm pix} > 75$ at different locations in the $T_{\rm eff}-\log g$ diagram. Most $T_{\rm eff}-\log g$ region where the RC stars concentrate has a contamination rate $< 5\%$. At the outer part, due to the lack of training set, the contamination rate is higher.}
\label{fig3}
\end{figure*}

We now construct a mapping from the normalized spectra, i.e., a vector with a dimensionality of the number of pixels, $N_{\rm pix}$, to the two asteroseismic values $\Delta \nu$ and $\Delta P$. This is a highly non-linear mapping that we construct through standard ``machine-learning'' approaches. We adopt a simple neural network that consists of three fully connected layers, each of which with $N_{\rm nodes} = 300$ hidden nodes, and the nodes are connected with a sigmoid activation function $\sigma$. We explored more elaborate networks, such as convolutional neural networks, but found them not to improve the classification, presumably because the mapping at hand is still relatively straightforward. Therefore, we keep the network simple for mathematical readability and the speed of training the network. Specifically, we presume that the asteroseismic $\Delta \nu$ and $\Delta P$ can be written as a function of the normalized flux ${\bf s}$, where
\begin{eqnarray}
(\Delta \nu, \Delta P) &\equiv& f({\bf s}) \\
= w_i^j &\sigma& \Bigg( w^k_j \sigma \bigg( w^l_k \sigma \Big( w^m_l s_m + b_{l} \Big) + b_k \bigg) + b_j \Bigg) + b_i,
\label{eq:the-payne}
\end{eqnarray}

\noindent
where $i \in \{1, 2\}$, $j, k, l \in \{ 1, \ldots, N_{\rm nodes}\}$ and $m \in \{ 1, \ldots, N_{\rm pix} \}$. We consider an $L1$ loss function to optimize the hyperparameter $\mathbf{p} \equiv ({\bf w}, {\bf b})$ to best describes the training set. In other words, we minimize $\sum_{i=1}^{N_{\rm train}} |(\Delta \nu, \Delta P)_i - f({\bf s_i}|{\bf p})|$, where ${\bf s}_i$ are the normalized fluxes of the $i$-th training spectra and $(\Delta \nu, \Delta P)_i$ are their corresponding asteroseismic value from \citet{vra16}. We adopt the python {\sc pytorch} package for this optimization. We did not consider the observed flux and asteroseismic parameter uncertainties, and adopted simply the mean values for the flux and the asteroseismic parameters. We explored whether the inclusion of flux uncertainties and 2MASS photometry as extra input in training significantly improved the asteroseismic parameter prediction, but found this not to be the case.

\begin{table*}
\begin{center}
\caption{Asteroseismic parameters added catalogs for LAMOST. Stars that have $\Delta P > 250\,$s are classified as RC stars.\\ This table is available in its entirety (with 413$,$472 rows) in machine-readable form.\label{table1}}
\begin{tabular}{lccccccccc}
\tableline \tableline
\\[-0.2cm]
Designation & RA  [$\deg$] & Dec  [$\deg$] & S/N & $\Delta P\,$[s] & $\Delta \nu\,[\mu$Hz] & Classification & $T_{\rm eff}\,$[K]\footnotemark[1]\footnotetext[1]{The stellar parameters values are adopted from LAMOST DR3.} & $\log g$\footnotemark[1] & [Fe/H]\footnotemark[1] \\[0.1cm]
\tableline
\\[-0.2cm]
J220430.94-011616.7 & 331.12894 & -1.27132 & 18.6 & 236.87 & 3.48 & -- & 4940 & 2.04 & -1.36 \\[0.1cm]
J220432.60+005112.5 & 331.13586 & 0.85349 & 199.5 & 76.99 & 5.02 & -- & 4725 & 2.69 & -0.45 \\[0.1cm]
J030142.28+001548.8 & 45.42620 & 0.26358 & 9.9 & 354.40 & 5.55 & RC\footnotemark[2] & 4901 & 2.53 & -0.57 \\[0.1cm]
J233420.50+332151.5 & 353.58542 & 33.36431 & 86.5 & 302.82 & 4.55 & RC\_Pristine\footnotemark[2]\footnotetext[2]{For LAMOST, we further distinguish RC stars with S/N$_{\rm pix} >75$ to be "RC\_Pristine", and RC stars with S/N$_{\rm pix} < 75$ to be "RC".} & 4937 & 2.47 & -0.46 \\[0.1cm]
$\cdots$ & $\cdots$ & $\cdots$ & $\cdots$ & $\cdots$ & $\cdots$ & $\cdots$ & $\cdots$ & $\cdots$ & $\cdots$ \\[0.1cm]
\tableline
\end{tabular}
\end{center}
\end{table*}

\begin{table*}
\begin{center}
\caption{Asteroseismic parameters added catalogs for APOGEE. Stars that have $\Delta P > 250\,$s are classified as RC stars.\\ This table is available in its entirety (with 80$,$944 rows) in machine-readable form.\label{table2}}
\begin{tabular}{lccccccccc}
\tableline \tableline
\\[-0.2cm]
Designation & RA  [$\deg$] & Dec  [$\deg$] & S/N & $\Delta P\,$[s] & $\Delta \nu\,[\mu$Hz] & Classification & $T_{\rm eff}\,$[K]\footnotemark[1]\footnotetext[1]{The stellar parameters values are adopted from APOGEE DR14.} & $\log g$\footnotemark[1] & [Fe/H]\footnotemark[1] \\[0.1cm]
\tableline
\\[-0.2cm]
2M00000211+6327470 & 0.00880 & 63.46308 & 122.9 & 233.62 & 3.33 & -- & 4694 & 2.42 & 0.02 \\[0.1cm]
2M00000446+5854329 & 0.01860 & 58.90915 & 148.5 & 313.75 & 3.82 & RC\_Pristine\footnotemark[2]\footnotetext[2]{For APOGEE, since most stars have S/N$_{\rm pix} >75$, we assume all RC stars to be "RC\_Pristine".} & 4756 & 2.37 & -0.02 \\[0.1cm]
2M00000535+1504343 & 0.02231 & 15.07621 & 184.9 & 84.17 & 14.81 & -- & 4914 & 3.22 & -0.05 \\[0.1cm]
2M00000866+7122144 & 0.03610 & 71.37069 & 495.8 & 82.83 & 4.34 & -- & 4685 & 2.55 & -0.04 \\[0.1cm]
$\cdots$ & $\cdots$ & $\cdots$ & $\cdots$ & $\cdots$ & $\cdots$ & $\cdots$ & $\cdots$ & $\cdots$ & $\cdots$ \\[0.1cm]
\tableline
\end{tabular}
\end{center}
\end{table*}

%
%
%
%
%
%

\section{Results and Discussion}

We now show the results of predicting asteroseismic parameters directly from APOGEE and LAMOST spectra. The right panels of Fig.~\ref{fig1} illustrate immediately that this works, as single-epoch spectra can predict $\Delta \nu$ and $\Delta P$ well, echoing \citet{haw18}. The left panels of Fig.~\ref{fig1} show the ``ground truth" for the LAMOST and APOGEE training samples, the distribution of asteroseismic $\Delta \nu$ and $\Delta P$ with its RGB and RC dichotomy. The right panels show the predictions for the same parameters, but now derived from the much larger test sample using our modelling of the LAMOST and APOGEE spectra; the distribution of these $\Delta \nu$ and $\Delta P$ predictions show the exact same morphology as the training set, attesting at least qualitatively the fidelity of the spectroscopic $\Delta \nu$ and $\Delta P$ estimates. Even though spectra probe only the photospheric properties of the stars, they ``know'' about the evolutionary states through subtle effects such as the [C/N] ratio. Classically, estimating $\Delta \nu$ and $\Delta P$ requires multi-epoch light curve from dedicated surveys, such as Kepler, and a careful ``boutique analysis'' \citep[e.g.][]{bed11}. Providing estimates of asteroseismic parameters from vastly abundant spectroscopic data is therefore valuable by itself. We note however that our "data-driven" predicted asteroseismic values are tied to the absolute scale outlined in \citet{vra16} and inherit any biases that the catalog might have.

But most relevant for the paper at hand, Fig.~\ref{fig1} illustrates that the spectroscopically estimated asteroseismic parameters, especially $\Delta P$, should allow us to cleanly separate the RC stars from the RGB stars \citep{bed11}. This is because RC stars have very different $\Delta P$ $(\sim 300\,$s) than their RGB cousins ($\sim 70\,$s), even they might share the same $T_{\rm eff}$ and $\log g$. The two modes in $\Delta P$ are clearly visible for both test sample in the right panels of Fig.\ref{fig1}. We adopt a conservative approach to identifying RC stars by choosing $\Delta P > 250\,$s, which minimizes the (false positive) contamination in the intermediate $\Delta P$ region. With this criterion, our method yields 179$,$286 RC stars from LAMOST, 58$,$343 of which are from LAMOST spectra with S/N$_{\rm pix} > 75$, and 36$,$908 RC stars from APOGEE. The results of our modelling are summarized in Table~\ref{table1} and Table~\ref{table2}.

In this study we consider both the primary RC stars (formed from lower mass stars) and the secondary RC stars (from more massive stars) together. But the bottom right panel of Fig.~\ref{fig1} suggests that it is also possible to classify primary and secondary RC stars using this method. At least for APOGEE, there is a visible bifurcation in $\Delta \nu$ for stars with $\Delta P \gtrsim 250\,$s. Stars with $\Delta \nu \gtrsim 6 \mu$Hz may be classified as secondary RC stars \citep{yan12}. However, due to the lack of secondary RC stars in the training sample, we do not attempt to separate the two groups. Nonetheless using $\Delta \nu \gtrsim 6 \mu$Hz as a selection criterion, we found that the ``contamination'' from secondary RC samples is pretty negligible and only consists of $\sim 2-4\%$ of our RC sample.

We emphasize that the training of the spectral model was done entirely independently for the APOGEE and the LAMOST data sets. Therefore, stars that were observed by both APOGEE and LAMOST constitutes excellent cross-validation set to quantify the contaminate rate, which is the focus of Fig.~\ref{fig2}. The right panel of Fig.~\ref{fig2} shows the $\Delta P$ spectroscopic estimates for all overlapping test stars (14$,$442 stars), showing good agreement between the two surveys. The standard deviations of $\Delta P$ and $\Delta \nu$ between these two surveys are 50$\,$s and 1$\,\mu$Hz, indicating that APOGEE and LAMOST can estimate asteroseismic parameters to such precision. However, due to the current small overlapping sample size between the asteroseismic and spectroscopic data, it is hard to estimate the exact contamination rate directly from the asteroseismic ``ground truth.'' This situation will improve with surveys such as TESS which will provide more stars with asteroseismic estimates in very near future. Nonetheless, in this study, we attempt to quantify the contamination rate from the APOGEE-LAMOST overlapping sample. We note that since APOGEE has higher resolution spectra and higher S/N (typically S/N$_{\rm pix} \simeq 200$), the contamination is mostly dominated by the LAMOST estimates. So we will primarily estimate the contamination for LAMOST below. We also tested the case where we set aside a part of APOGEE spectroscopic-asteroseismic sample as a cross-validation set. The cross-validation set estimates that the contamination rate for APOGEE RC sample is about $2-3\%$, for the range of S/N obtained in APOGEE, in agreement with \citet{haw18}.

The left panel of Fig.~\ref{fig2} shows the contamination rate for the RC classification for the LAMOST sample, as a function of S/N. The contamination rate is estimated from the differences in the $\Delta P$ estimates from APOGEE and LAMOST, for the stars that overlap among the two samples. We take it to be the fraction of stars that have $\Delta P_{\rm LAMOST} > 250\,$s, but with $\Delta P_{\rm APOGEE} < 250\,$s (upper, pessimistic contamination limit) or $\Delta P_{\rm APOGEE} < 200\,$s (lower, optimistic limit). The color-coding indicates the normalized number density of the LAMOST test data. The panel illustrates that the contamination rate is a strong function of S/N: for poor S/N, low-resolution LAMOST spectra can incur a contamination rate of $10-20\%$, a similar rate as previous approaches \citep{bov14,wan15}. But when analyzing the LAMOST data with higher S/N, our approach does very well: at S/N$_{\rm pix}= 50$, the contamination rate is about $5-10\%$ and at S/N$_{\rm pix}\,=100$, the contamination is only $\sim 3\%$. The low contamination rate is perhaps not surprising as we exploit the full spectral information, beyond the stellar parameters that were previously explored. As shown in the $y$-axis, the overall contamination rate is about $9\%$, when integrating over the full range of S/N. This is two times lower than previous approaches for LAMOST, yielding 179$,$286 presumed RC stars. Since the primary goal of this study is to present a pristine catalog, in the following, we only select stars with S/N$_{\rm pix} > 75$. This criterion leaves 58$,$343 LAMOST RC stars, with a contamination rate of only $3\%$. When combined with the APOGEE non-duplicating sample, we have a total 92$,$249 ``pristine'' RC stars. Here we publish both our classifications, as well as the spectroscopically inferred asteroseismic parameters that permit independent classifications.

While the S/N cut severely impacts the completeness of our RC pristine catalog, as shown in the top $x$-axis, we emphasize that our approach is primarily limited only by the spectra quality, not the astrophysical degeneracy in stellar parameters between RC and RGB stars. We caution however that the completeness stated only aims to serve a rough guide.  Due to the lack of spectroscopic-asteroseismic training ``ground-truth", here we adopt a somewhat ad-hoc definition for completeness. We define the completeness to be the fraction of the number of stars with $\Delta P > 250\,$s to the total number of stars with $\Delta P > 200\,$s, assuming that the number of RC stars that get wrongly classified to have $\Delta P < 200\,$s is compensated by the number RGB stars classified with $\Delta P > 200\,$s.

Another advantage of this method is that it is incredibly efficient and does not require deriving stellar parameters beforehand. Training the empirical relation takes $< 5$ GPU minutes, and inferring the asteroseismic catalog for all 413$,$472 LAMOST stars only takes $\sim 1$ CPU minute. Therefore, one can imagine an efficient survey strategy would be collecting low S/N data in the first pass, and stars with high estimated $\Delta P$ will then be followed up to improve the S/N and provide a more accurate classification.

\begin{figure}
\centering
\includegraphics[width=0.5\textwidth]{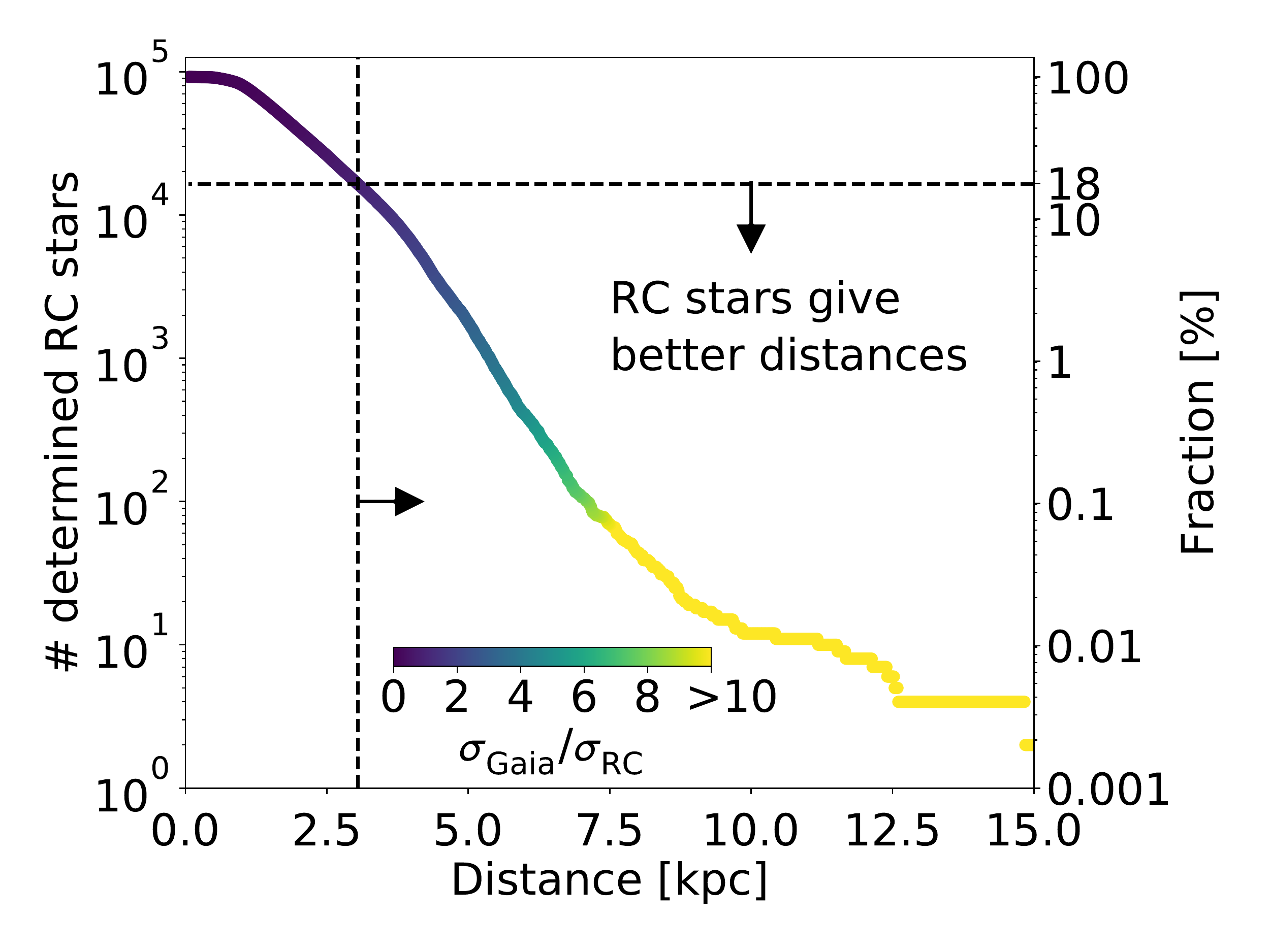}
\caption{RC stars have excellent distance estimates even in the Gaia era. We show the cumulative distance distribution of the combined pristine RC sample from APOGEE and LAMOST detected in this study, excluding duplicates and low S/N (median S/N$_{\rm pix} < 75$) LAMOST objects. The color-coding shows the ratio of the estimated Gaia end-of-mission uncertainties to the typical distance uncertainties for RC stars. While Gaia provides better distances for stars that are within 3 kpc from the Sun, the precision of parallax distances is limited for stars located at a more considerable distance. RC stars, on the other hand, is only limited by photometric uncertainties and can provide better distances for a significant fraction of the Milky Way. About $18\%$ (16$,$411 stars) of this pristine sample in this study will continue to provide better distances even with the Gaia final catalog.}
\label{fig4}
\end{figure}

Fig.~\ref{fig3} shows that our determined RC stars agree well with stellar evolution models. The left and middle panels of Fig.~\ref{fig3} show the $T_{\rm eff}$ and $\log g$ of the RC stars (without S/N cut). The black lines indicate the stellar evolution models from the MIST isochrones \citep{cho16} at 4 Gyr. The grey background illustrates the number density of all test stars and the blue symbols the training sample. These panels also demonstrate that, for the high-resolution APOGEE sample, since the $T_{\rm eff} - \log g$ estimates are sufficiently accurate, the classical $T_{\rm eff} - \log g$ method \citep[e.g.,][]{bov14} can do relative well. On the other hand, the overlap in $T_{\rm eff}$ and $\log g$ is not trivial for the low-resolution LAMOST sample, and the method presented in this study excels by exploring spectral information beyond the stellar parameters to achieve a low contamination rate. To further illustrates this point, the right panel shows the average contamination rate for different $T_{\rm eff}$ and $\log g$. The background shows the number density of the APOGEE-LAMOST overlapping sample. We only show bins in which the overlapping sample has more than 5 RC stars. As shown, for LAMOST data with S/N$_{\rm pix} >75$, the contamination rate is $< 5\%$ even for regions where RC stellar parameters directly overlap with the RGB stars. Nonetheless, the panel also shows that in the part where there is a lack of training data, the contamination can still be high and can only improve with a more extensive training sample collected in the near future.

Constructing this pristine catalog of RC stars as standard candles across the Milky Way begs the question of whether it will retain value in light of Gaia data, which provide parallxes for a billion stars. In Fig.~\ref{fig4} illustrates that knowing that a star is an RC stars provides better distance estimates for the most part of the Milky Way. Plotted in Fig.~\ref{fig4} is the cumulative distribution of the number of pristine RC stars determined in this study as a function of heliocentric distances. We estimate the heliocentric distances by extrapolating the linear relation between $\log (d)$ and 2MASS $K$ band magnitude determined in \citet{haw17}, where $d$ is the heliocentric distances. We emphasize that the distances are only rough estimates as we do not attempt to correct for color, stellar age, and extinction. The full distances and extinctions will be carried out in a future study following the hierarchical Bayesian method developed in \citet{haw17}. We crossmatch our sample with Gaia DR1 to obtain the Gaia $G$ band magnitude and calculate the end-of-mission Gaia parallax uncertainties following the estimates from \citet{bru14}. This is a conservative limit since the coming Gaia DR2 parallax is about 2 times worse than the final catalog.\footnote{https://www.cosmos.esa.int/web/gaia/dr2} As for the RC stars, we assume that the distances uncertainties are about $6\%$ \citep[e.g.,][]{bov14, haw17}. The fact that RC stars are better distance indicators for distant stars is expected: RC stars are brighter than $G < 15$ within 10 kpc from the Sun, and are therefore not severely limited by the photometric uncertainties, unlike astrometric measurements which quickly degrades with distances. We also note that while the distances in Fig.~\ref{fig4} truncates at 20$\,$kpc, this is due to the S/N cut which biases against distant stars. The full catalog in this study with no S/N cut has a sample size that is 5 times larger and reaches $\sim 30\,$kpc. Furthermore, the current LAMOST iDR5 (not publicly available) catalog is 2 times larger. So revisiting some of the low LAMOST S/N RC candidates can readily provide an RC sample catalog that is about $\sim 500$,$000$ RC stars with $3 \%$ contamination and $\sim 3\%$ distances out to $\sim 30 \,$kpc.

%
%
%
%
%
%

\section{Summary and future outlook}

In this study, we present a catalog of $210$,$371$ RC stars with contamination of $9\%$, derived from the APOGEE and LAMOST surveys. Among this sample, if we only consider spectra with S/N$_{\rm pix}>75$, we have a pristine subsample of 92$,$249 RC clumps stars with $3\%$ contamination, which is among the largest RC catalogs but with the least contamination. 

We show that while single epoch spectra only probe the photospheric properties of stars, a data-driven model can be established to predict asteroseismic parameters, in particular the frequency separation of stellar interior $p$-modes $\Delta \nu$ and the mixed mode period spacing $\Delta P$ consistent with \cite{haw18}, here we present the spectral inferred $\Delta \nu$ and $\Delta P$ for about 500$,$000 stars. More importantly, such a data-driven model exploits more spectral information beyond the stellar parameters ($T_{\rm eff}, \log g, [{\rm M}/{\rm H}])$. As a result, the method yields a more pristine RC catalog through their inferred asteroseismic parameters, especially low-resolution LAMOST spectra because low-resolution spectra are more limited in the precision of $T_{\rm eff} - \log g$ which hamper the ability to separate out RC and RGB stars using traditional methods.

The RC catalog presented in this work provides an excellent opportunity to map the Galaxy in a complimentary way to Gaia particularly for distant stars. But more importantly, with surveys such as LAMOST, DESI, and TESS which will provide many more spectroscopic-asteroseismic samples and low-resolution spectra from a significant fraction of stars in the Milky Way, the method presented in this study provides a straightforward way to select a pristine catalog of RC stars without the need of inferring stellar parameters. With low-resolution surveys such as LAMOST and DESI, we should be able to find an extensive and pristine sample of RC stars of about 500$,$000 RC stars out to 30$,$kpc. A dedicated effort can then be planned to follow up these RC stars to perform the ultimate Galactic cryptography of the Milky Way.

%
%
%
%
%
%

\acknowledgments

YST is grateful to be supported by the Martin A. and Helen Chooljian Membership from the Institute for Advanced Study in Princeton, the joint Carnegie-Princeton Fellowship from Princeton University and Carnegie Observatories and the Australian Research Council Discovery Program DP160103747. KH is funded by the Simons Foundation Society of Fellows and the Flatiron Institute Center for Computational Astrophysics in New York City. HWR's research contribution is supported by the European Research Council under the European Union's Seventh Framework Programme (FP 7) ERC Grant Agreement n.$,$[321035] and by the DFG's SFB-881 (A3) Program. This project was developed in part at the 2017 Heidelberg Gaia Sprint, hosted by the Max-Planck-Institut for Astronomie, Heidelberg. This study makes use of the publicly released data from LAMOST DR3 and APOGEE DR14. The Guoshoujing Telescope (the Large Sky Area Multi-Object Fiber Spectroscopic Telescope LAMOST) is a National Major Scientific Project built by the Chinese Academy of Sciences. Funding for the project has been provided by the National Development and Reform Commission. LAMOST is operated and managed by the National Astronomical Observatories, Chinese Academy of Sciences. The Sloan Digital Sky Survey IV is funded by the Alfred P. Sloan Foundation, the U.S. Department of Energy Office of Science, and the Participating Institutions and acknowledges support and resources from the Center for High- Performance Computing at the University of Utah.

%
%
%
%
%
%

\end{CJK*}

\vspace{1cm}

\bibliography{biblio.bib}

\end{document}